\documentclass[12pt]{article}
\newcommand{\np}[3]{{ Nucl.Phys.} {\bf #1} (#2) #3}
\newcommand{\pr}[3]{{ Phys.Rev.} {\bf #1} (#2) #3}

\newcommand{\pl}[3]{{ Phys.Lett.} {\bf #1} (#2) #3}

\newcommand{\hepth}[1]{\textsf{hep-th/#1}}

\newcommand{\eqref}[1]{(\ref{#1})}              % (1.16), not 1.16

\newcommand{\gymsq}{\ensuremath{g^{2}_{\mathrm{YM}\ }}}    % g_ym^2

\newcommand{\tr}{\mbox{tr}} 	 			% trace
\newcommand{\Tr}{\mbox{Tr}}				% Trace
\newcommand{\STr}{\mbox{STr}}				% STr
\newcommand{\diag}{\mbox{diag}\,}			% diagonal matrix

\renewcommand{\thefootnote}{\fnsymbol{footnote}}

\begin{document}

\begin{titlepage}
\begin{flushright}
  Imperial/TP/97-98/58  \\
  hep-th/9806186  \\
\end{flushright}
\vspace{1.5cm}

\begin{center}
{\LARGE   
  Probing a D6  +  D0 state with D6-branes:\\
  \vspace{0.2cm} 
   SYM - Supergravity correspondence \\
  \vspace{0.3cm} 
  at subleading level 
  \footnote{Work supported by FCT - Programa PRAXIS XXI, 
       under contract BD/9347/96}
   }\\[.2cm]
\vspace{2cm}
{\large Jo\~{a}o Branco\footnote{\ E-mail: J.Branco@ic.ac.uk}} \\
\vspace{18pt}
{\it Theoretical Physics Group, \\
	Blackett Laboratory, \\
	Imperial College,\\ 
	London SW7 2BZ, U.K.}
\end{center}
\vskip 3 cm

\begin{abstract}
  We probe a non-supersymmetric D6 + D0 state with D6-branes
  and find agreement at subleading order between the supergravity and 
  super Yang-Mills description of the long-distance, low-velocity 
  interaction.
\end{abstract}

\end{titlepage}
\setcounter{page}{1}
\renewcommand{\thefootnote}{\arabic{footnote}}
\setcounter{footnote}{0}

%%%%%%%%%%%%%%%%%%%%%%%%%%%%%%%%%%%%%%%%%%%%%%%%%%%%%%%%%%%%%%%%%%%%%%

\section{Introduction}

During the last year and a half there was an 
intensive study of the correspondence 
between the Supergravity and the Super Yang-Mills (SYM) 
descriptions 
of the long distance interactions between D$p$-branes and their bound 
states \cite{bac9511,lif9610,bfss9610,che-tse9705,
mal9709, kab-tay9712}
\footnote{This is by no means an exhaustive list. For more 
complete references see e.g. \cite{kab-tay9712}.}. 
Most of this work considered only the leading $F^4$ terms in the 
1-loop SYM effective action, for which there is a known 
expression for a general gauge field background. 
For the specific case of D0-brane \-- D0-brane interaction, 
the subleading terms were directly computed in SYM theory, 
and once again agreement 
was found with the supergravity  
calculations \cite{bec-bec9705,bbpt9706}.

This led to the conjecture \cite{che-tse9709,mal9709}
(see also~\cite{bgl9712,kes-kra9709}) 
that the leading 
part of the $L$-loop term in the SYM effective action in $D=p+1$ 
dimensions has a universal $F^{2L+2}/M^{(7-p)L}$ structure, and 
that this term, when computed for a SYM background representing 
a configuration of interacting branes and in the large $N$ limit, 
should reproduce the 
$1/r^{(7-p)L}$ term in the corresponding long-distance 
supergravity potential.
The authors of \cite{che-tse9709} then proposed a specific form 
for the $F^6$ terms in the 2-loop effective action, and 
proceeded to show that it reproduces the subleading supergravity 
potentials for a variety of configurations.
However, only configurations that still preserved some 
($1/2$, $1/4$ or $1/8$) supersymmetry were considered. 
The proposed $F^6$ term has not yet been tested 
for a {\em non-supersymmetric} configuration.

The existence of extreme but non-supersymmetric black holes was 
shown in~\cite{khu-ort9512}, where a supergravity 
background that 
after compactification to 10 
dimensions describes a dyonic non-supersymmetric 
black hole was found. A different 
compactification scheme was used in~\cite{she9705} 
to obtain an extremal non-supersymmetric solution 
to low energy type 
IIA string theory carrying D0- and D6-brane 
charges. This solution, for non-zero values of both 
charges,  breaks all supersymmetries.
The scattering of D0- and D6-branes from these states was studied 
in \cite{kes-kra9706,pie9707}, and considering a more 
general supergravity background in~\cite{bisy9711} (applying the SYM 
results of~\cite{mal9705,mal9709}). Once again 
agreement was found between the leading terms in supergravity  
and the SYM 
expressions at 1-loop\footnote{
A very similar supergravity background was found in~\cite{dha-man9803}, 
describing a 4-dimensional black hole carrying D0- and D6-brane 
charges. The black hole was probed with 
D0-branes both in the supergravity and M(atrix) theory formalism, 
with the 
by now expected agreement at 1-loop.}.

Our aim in the present paper is to  
extend the supergravity calculation~\cite{bisy9711} 
of the source-probe potential 
to subleading terms in $1/r$, and compare these with those
obtained using the ansatz of~\cite{che-tse9709} for the 2-loop SYM 
effective action.

%%%%%%%%%%%%%%%%%%%%%%%%%%%%%%%%%%%%%%%%%%%%%%%%%%%%%%%%%%%%%%%%%%%%%%

\section{Supergravity description}

In this section, we use the same normalisation as in~\cite{bisy9711}, 
$\alpha^{\prime}=1$.

In the supergravity calculation, one considers a probe moving in 
the background created by a source. Both the probe and source can 
be clusters of branes or of bound states of branes.

The relevant part of the action of a D$p$-brane 
(containing  no lower dimensional branes) 
in a supergravity background is

\begin{equation}
  S = T_{p}\int d^{p+1} x \; \left[e^{-\phi}
    \sqrt{-\det (G_{\mu\nu}+B_{\mu\nu})} + 
  \frac{1}{(p+1)!}\epsilon^{\mu_1\mu_2\ldots\mu_{p+1}}
  C_{\mu_1\mu_2\ldots\mu_{p+1}}\right],
 \label{pbaction}
\end{equation}

where $G_{\mu\nu}$, $B_{\mu\nu}$ and $C_{p+1}$
are the pullbacks to the 
world-volume of the background 
10D metric, 2-form and  
R-R $(p+1)$-form (for  
`electric' branes) or the Hodge dual of the 
R-R $(7-p)$-form (for `magnetic' branes), respectively,  
the brane tension is \cite{pol9611}

\begin{equation}
  T_{p}=n_p g_s^{-1} (2\pi)^{(1-p)/2}T^{(p+1)/2}\;,
  \label{tension}
\end{equation}

with $g_s$ the string coupling constant, 
and the string tension is $T=(2\pi \alpha ')^{-1}$.

%%%%%%%%%%%%%%%%%%%%%%%%%%%%%%%%%%%%%%%%%%%%%%%%%%%%%%%%%%%%%%%%%%%%%

\subsection{The supergravity background}

In~\cite{bisy9711}, to which we refer the reader for 
further details, a solution to the low energy  
effective action of type IIA string theory,

\begin{equation}
  S = \frac{1}{(2\pi)^7 g_s^2}\int d^{10} x \: \sqrt{-g_{10}}
  \left[e^{-2\phi}\left(R_{10}+4(\nabla\phi)^2\right)
		-\frac{1}{4}F_{\mu\nu}^2\right],
\end{equation}

was found 
describing a system 
with D0- and D6-branes.

The spherically symmetric, time-independent solutions 
are parametrised by the mass ($M$), electric charge ($Q$) 
and magnetic charge ($P$). The dilaton charge ($\Sigma$)
is related to these by

\begin{equation}
  \frac{8}{3}\Sigma = \frac{Q^2}{\Sigma + \sqrt{3}M} + 
       		    \frac{P^2}{\Sigma - \sqrt{3}M}\;.
	\label{dilcharge}
\end{equation}

The explicit form of the fields is 

\begin{eqnarray}
  e^{4\phi/3} & = &  \frac{B}{A}\;, \label{dil} \\
  A_{\mu} dx^{\mu} & = & \frac{Q}{B}(r-\Sigma)dt + P \cos \theta d\phi\;, 
	\label{1form} \\
  g_{\mu\nu}dx^{\mu}dx^{\nu} & = & -\frac{F}{\sqrt{AB}}dt^2 + 
  \sqrt{\frac{B}{A}}(dx_1^2 + \cdots + dx_6^2) + 
  \frac{\sqrt{AB}}{F}dr^2 \nonumber \\
  & & + \sqrt{AB}(d\theta^2 + \sin^2 \theta d\phi^2)\;,
\end{eqnarray}

where
\begin{eqnarray}
  F & = & (r-r_{+})(r-r_{-})\;, \nonumber \\
  A & = & (r-r_{A+})(r-r_{A-})\;, \\
  B & = & (r-r_{B+})(r-r_{B-})\;, \nonumber
\end{eqnarray}

and 
\begin{eqnarray}
  r_{\pm} & = & M \pm \sqrt{M^2 + \Sigma^2 - 
	\frac{P^2}{4}-\frac{Q^2}{4}}\;, \nonumber \\
  r_{A\pm} & = & \frac{\Sigma}{\sqrt{3}}\pm
	\sqrt{\frac{P^2\Sigma/2}{\Sigma-\sqrt{3}M}}\;, \\
  r_{B\pm} & = & -\frac{\Sigma}{\sqrt{3}}\pm
	\sqrt{\frac{Q^2\Sigma/2}{\Sigma+\sqrt{3}M}}\;. \nonumber
\end{eqnarray}

The extremality condition, $r_+ = r_-$\,, is equivalent to

\begin{equation}
  M^2 + \Sigma^2 - \frac{P^2}{4}-\frac{Q^2}{4} = 0\;.
	\label{extremal}
\end{equation}

When imposing this condition it 
turns out to be convenient to perform the coordinate change 
$r \rightarrow r^{\prime} =  r - M$. The extremal solution is then written as 
(we drop the prime on $r$)

\begin{equation}
  ds^2= -f_1(r)dt^2 + f_2(r)dx_i dx_i + f_1^{-1}(r)(dr^2+r^2d\Omega^2)\;,
	\label{06back}
\end{equation}

where

\begin{equation}
  f_1(r) = \frac{r^2}{\sqrt{AB}}\;,\;\;\;\;\;\;\; 
	f_2(r) = \sqrt{\frac{B}{A}}\;.
\end{equation}

It is easy to see that the pure magnetic solution

\begin{equation}
  P=4M\;, \;\;\;\; Q=0\;,\; \;\;\;\Sigma=-\sqrt{3}M\;,
\end{equation}

solves the extremality condition~\eqref{extremal} 
and describes a D6-brane.

At the same time, the pure electric solution

\begin{equation}
  P=0\;, \;\;\;\;Q=4M\;, \;\;\;\;\Sigma=\sqrt{3}M\;,
\end{equation}

 also solves~\eqref{extremal} and describes D0-branes smeared over a 6-torus.

We expect that solutions with both electric and magnetic charge 
interpolate between these and so describe a system with 
both D0- and D6-branes.

It turns out (\cite{bisy9711} and references therein) 
 that the mass, electric and magnetic charge are related 
to the number of branes as 

\begin{eqnarray}
  M & = & \frac{g_s N_6}{8}\;, \nonumber \\
  P & = & \frac{g_s N_6}{2}\;, \label{MPQ} \\
  Q & = & \frac{g_s N_0(2\pi)^2}{2V_6}\;. \nonumber
\end{eqnarray}

We can now describe a system with a large number of D6-branes and 
a relatively 
small number of D0-branes, i.e., $P \gg Q $. To do so we consider 
some fixed $M$, and move away from the pure magnetic solution, 
by taking

\begin{equation}
  \Sigma = M(-\sqrt{3}+\epsilon)\;,\;\;\;\;\;\;\;\; \epsilon \ll 1\;.
\end{equation}

Note that by doing this we are moving away from a $1/2$ 
supersymmetric solution, a D6-brane, 
towards one that breaks all supersymmetries.

The dilaton charge equation~\eqref{dilcharge} 
and the extremality condition~\eqref{extremal} 
allow us to determine the electric and magnetic charge of this 
system as 

\begin{eqnarray}
  P  & = & \frac{1}{3}\sqrt{2}\sqrt{72M^2-36\sqrt{3}\epsilon M^2
	+18\epsilon^2 M^2-\sqrt{3}\epsilon^3 M^2}\;, \\
  Q & = & \frac{\sqrt{2}\epsilon^{3/2}M}{3^{3/4}}\;,
\end{eqnarray}

or, to leading order in $\epsilon$,

\begin{eqnarray}
  P & = & 4M-\sqrt{3}M\epsilon+\frac{M\epsilon^2}{8} + O(\epsilon^3)\;, 
		\label{P} \\
  Q & = & \frac{\sqrt{2}\epsilon^{3/2}M}{3^{3/4}}\;. \label{Q}
\end{eqnarray}

For these values of the parameters our solution is an extremal
(in the sense of~\eqref{extremal}), 
near-supersymmetric one, with the parameter 
$\epsilon$ measuring deviation from 
supersymmetry.

%%%%%%%%%%%%%%%%%%%%%%%%%%%%%%%%%%%%%%%%%%%%%%%%%%%%%%%%%%%%%%%%%%%%%

\subsection{Supergravity calculation}

We now consider a D6-brane probe moving in this background. 
We take the D6 probe to be parallel to the D6 
source, and assume the static gauge, i.e., that the worldvolume 
coordinates of the D6 probe are the same as those of the source, and that 
the transverse coordinates do not depend on the spatial 
worldvolume coordinates.

With these assumptions, the long distance, low velocity
action of the D6 probe can be obtained 
from~\eqref{pbaction} by plugging in the 
background~\eqref{dil}, \eqref{1form}, \eqref{06back} for the values of 
the parameters~\eqref{P}, \eqref{Q} and expanding 
in $1/r$ and $v$, with $r$ the transverse distance between branes and 
$v$ the transverse velocity.

We obtain 
\footnote{
We have dropped here a term 
linear in $v$, since we will not 
compare it with the SYM results. A term of this type  
was recently discussed in~\cite{bvflrs9805}  for the potential between 
a D0- and a D6-brane in relative motion.}

\begin{eqnarray}
  S  & = & \frac{n_6}{g_s (2\pi)^6} V_6 \times \nonumber \\
  && \int dt 
  \left[ 1+\frac{\epsilon^2 M}{8r} - 
   \left(\frac{1}{2}+\frac{3\epsilon^2 M^2}{4 r^2}+
   \frac{\sqrt{3}\epsilon M}{2r} \right)v^2 \right. \nonumber \\
  && \;\;\;\;\;\;\;\;\;\;
   -\left(\frac{1}{8}+\frac{\sqrt{3}\epsilon M^2}{r^2}-
   \frac{\epsilon^2 M^2}{16 r^2}+\frac{M}{2r}+
   \frac{\sqrt{3}\epsilon M}{8r}\right)v^4 \nonumber \\
  && \;\;\;\;\;\;\;\;\;\;
   \left. -\left(\frac{1}{16}+\frac{M^2}{r^2}+
   \frac{\sqrt{3}\epsilon M^2}{r^2} -
   \frac{5\epsilon^2 M^2}{32r^2}+\frac{M}{2r}+
   \frac{\sqrt{3}\epsilon M}{16r}\right) v^6\right],
	\label{Ssugra06}
\end{eqnarray}

where $n_6$ is the number of D6 branes in the probe.

Note that in the limit $\epsilon \rightarrow 0$ the static term 
of the potential vanishes, and the corrections to the energy start 
only at $v^4$. This is what we should expect~\cite{tse9609}, 
since in this limit our 
background reduces just to a collection of overlapping 
D6-branes, 
and so the full system is just 
D6-branes parallel to D6-branes, which is a BPS configuration.

%%%%%%%%%%%%%%%%%%%%%%%%%%%%%%%%%%%%%%%%%%%%%%%%%%%%%%%%%%%%%%%%%%%%%%

\section{SYM description}

In this section, we use the normalisation of~\cite{che-tse9709}, 
$T=1$.

The low energy dynamics of 
$N$ D-branes is described by the dimensional reduction to 
$p+1$ dimensions of $\mathcal{N}=1$  
SYM in 10 dimensions with $U(N)$ gauge symmetry
(for a recent review and references see~\cite{tay9801}).

In \cite{che-tse9709,che-tse9801} 
it was argued that the sum of leading 
large $N$ IR contributions to the effective action 
can be written as 

\begin{equation}
  \Gamma = \sum_{L=1}^{\infty} \Gamma^{(L)} =    
   \frac{1}{2}\sum_{L=1}^{\infty} \int d^{p+1}x \left(\frac{a_{p}}
   {M^{7-p}}\right)^L (\gymsq)^{L-1}\hat{C}_{2L+2}(F)\;,
   \label{effsym}
\end{equation}

where $F$ is a background field, the coefficients 
$a_p$ are given by ($T=1$)

\begin{displaymath}
  a_{p} = 2^{2-p}\pi^{-(p+1)/2}\Gamma\left(\frac{7-p}{2}\right),
\end{displaymath}

and the coefficients $\hat{C}_{2L+2}$ are polynomials of $F$.
The only term which was explicitly computed is 
the 1-loop one, with the result 

\begin{equation}
  \hat{C}_4 = \STr\, C_4\;,
	\label{hc4}
\end{equation}

where STr is the symmetrised trace in the {\em adjoint} 
representation, and $C_4$ is as given below.

For the corresponding 2-loop term the authors 
of~\cite{che-tse9709} have proposed the ansatz

\begin{equation}
  \hat{C}_6 = \widehat{\STr}\, C_6\;,
   \label{hc6}
\end{equation}

where $\widehat{\STr}$ is a modified symmetrised trace, 

\begin{eqnarray}
\widehat{\STr}\,(X_{i_1}\ldots X_{i_6})& = & 
    2N\tr(X_{(i_1}\ldots X_{i_6)})+
	60\,\tr(X_{(i_1}\ldots X_{i_4})\tr(X_{i_5}X_{i_6)}) \nonumber \\
&& -50\,\tr(X_{(i_1}\ldots X_{i_3})\tr(X_{i_4}\ldots X_{i_6)}) 
	\nonumber \\
&&  -30 N^{-1} \tr(X_{(i_1}X_{i_2})\tr(X_{i_3}X_{i_4})
	\tr(X_{i_5}X_{i_6)})\;.
\end{eqnarray}

The coefficients $C_4$ and $C_6$ 
are the same as the polynomials appearing in the expansion of the 
abelian Born-Infeld action ($T=1$),

\begin{equation}
  \sqrt{-\det(\eta_{\mu\nu}+F_{\mu\nu})}= \sum_{n=0}^{\infty}
	C_{2n}(F)\;,
\end{equation}

with
\begin{eqnarray}
  & C_{0} = 1\;,\;\;\;\; C_{2}=-\frac{1}{4}F^{2}\;,\;\;\;\; 
	C_{4} = -\frac{1}{8}
  \left[F^{4}-\frac{1}{4}(F^{2})^{2}\right]\;,  & \nonumber \\
  & C_{6} = -\frac{1}{12}
   \left[F^{6}-\frac{3}{8}F^{4}F^{2}+\frac{1}{32}(F^2)^3\right]\: , \: 
   \ldots &  \label{BIc}
\end{eqnarray}

where $F^{k}$ is the trace of the matrix product over Lorentz 
indices,

\begin{displaymath}
  F^{2}=F_{\mu\nu}F^{\nu\mu}\;, \;\;\;\ldots \; ,\;\;\;
  F^{2k}=F_{\mu_1\mu_2}F^{\mu_2\mu_3} \ldots 
  F_{\mu_{2k-1}\mu_{2k}}F^{\mu_{2k}\mu_1}\;.
\end{displaymath}

\subsection{The SYM background}

A SYM background describing a system with D0- and D6-branes but no 
D2- nor D4-branes was found in~\cite{tay9705}.

Let
\footnote{We use the superscript `$S$' for `source', i.e., the 
background describing just the D0+D6 system, in order 
to avoid confusion 
with the full background~\eqref{F}, \eqref{Ji}.}
\begin{equation}
  F^{S}_{12} = F_0 J^{S}_1\;,\;\;\;\; F^{S}_{34} = F_0 J^{S}_2\;,\;\;\;\; 
  F^{S}_{56}=F_0 J^{S}_3\;, 
\end{equation}

where $F_0$ is an arbitrary constant, 
the $J_i^{S}$'s are $N_6 \times N_6$ block-diagonal matrices 
built out of $\frac{1}{4} N_6$ copies of $\mu_i$, 
$N_6$ a multiple of four,

\begin{equation}
  J^{S}_i = \diag(\mu_i, \ldots, \mu_i)\;, 
		\;\;\;\; i=1,2,3,
	 \label{Jsi}
\end{equation}

and $\mu_i$ are the $su(4)$ Cartan subalgebra matrices
\begin{eqnarray}
  \mu_1 & = & \diag(1,1,-1,-1)\;, \nonumber \\
  \mu_2 & = & \diag(1,-1,-1,1)\;, \\
  \mu_3 & = & \diag(1,-1,1,-1)\;. \nonumber 
\end{eqnarray}

They have the properties~\cite{tay9705}
\begin{eqnarray}
 & & \;\;\tr\, (\mu_i^2)  =   4\;, \nonumber  \\
 & & \left. \begin{array}{l}
   \tr\,({\mu_i})  =  0\;,\\
   \mu_i \mu_j  =  |\epsilon_{ijk}| \mu_k\,, \;\; i \neq j\, ,\\
   \end{array} \right\}
   \Rightarrow \tr\,(\mu_i \mu_j) = 0\, ,\;\; i\neq j\;.
\end{eqnarray}

Using these we easily evaluate the D0-, D2- and D4-brane charges 
induced by the flux of the background in the worldvolume,

\begin{eqnarray}
  N_4 & \propto & \int d^2 x \;\tr\,(F^{S}) = 0\;, \nonumber \\
  N_2 & \propto & \int d^4 x \;\tr\,(F^{S}\wedge F^{S}) = 0\;, \nonumber \\
  N_0 & = & \frac{T^3}{6(2\pi)^3} 
	\int d^6x\;\tr\,(F^{S}\wedge F^{S}\wedge F^{S}) \nonumber \\
     & = & 	\frac{F_0^3 N_6 V_6 T^3}{(2\pi)^3}\;, \label{N0}
\end{eqnarray}

so that, as required, there are no D2- or D4-brane charges, 
but only D0- 
and D6-brane ones. It can easily be seen~\cite{tay9705} that, for 
$F_0 \neq 0$, this is a 
non-supersymmetric state. This state is also not a bound state, 
since 
its energy is larger than that of the separated constituents. 
It is expected nevertheless that it should represent 
a meta-stable state.

We will probe this system with D6-branes. We consider a D6-brane probe
parallel to the D6 source, and for simplicity consider them 
aligned with the coordinate axis $(0)123456$. The probe will have 
transverse velocity $v$, and we will take its direction to be along 
the $9$ axis.

The SYM background describing the full system (source + probe)  
is (see e.g.~\cite{che-tse9709}),

\begin{equation}
  F_{12} = F_0 J_1\;,\;\;\; F_{34} = F_0 J_2\;,\;\;\; F_{56}=F_0 J_3\;, 
  \;\;\;F_{09} = v J_0\;,
          \label{F}
\end{equation}

where the $J$'s are block-diagonal matrices, 

\begin{eqnarray}
  J_i & = & \diag\!\left(0_{n_6 \times n_6},(J^{S}_i)_{N_6\times N_6}\right)
           ,\;\;\;\;i=1,2,3,
	 \label{Ji} \\
   J_{0} & = & \frac{1}{N} \diag\!\left(N_6 I_{n_6 \times n_6} ,  
	-n_6 I_{N_6 \times N_6}\right)\;, \;\;\;\;N \equiv n_6 + N_6\;.
	  \label{J0}
\end{eqnarray}

\subsection{SYM calculation}

Note that the $F$ matrices in~\eqref{F} 
are in the {\em fundamental} 
representation 
of $su(N)$, and the trace 
in~\eqref{hc4},~\eqref{hc6} is in the adjoint representation. 
We need therefore to know how to relate traces in the two 
representations. To that effect, we collect some useful formulas in 
Appendix A.

We start by evaluating the $F^4$. Plugging the SYM 
background~\eqref{Ji},\eqref{J0} into~\eqref{BIc} 
we get

\begin{eqnarray}
  C_4 & = & -\frac{1}{8}\left[F_0^4(J_1^4 + J_2^4 + J_3^4)
	- 2 F_0^4(J_1^2 J_2^2 + J_1^2 J_3^2 + J_2^2J_3^2) 
		\right. \nonumber \\
       & & \left.\;\;\;\;\;\;\;
		+\,2 F_0^2 (J_1^2 + J_2^2 + J_3^2)J_0^2 v^2 
			+ J_0^4 v^4\right],
\end{eqnarray}

from which, using the trace results of Appendix A,

\begin{equation}
  \hat{C}_4 = \frac{1}{4}n_6 N_6 (3 F_0^4 - 6 F_0^2 v^2 - v^4)\;.
\end{equation}

From~\eqref{effsym} we read the expression for  
the 1-loop effective action, 

\begin{eqnarray}
  \Gamma^{(1)} & = &  \frac{a_6}{2M}\int d^7x\;\hat{C}_4(F)
	\nonumber \\
   & = & \frac{n_6 N_6}{16 (2\pi)^3 r}V_6 
	\int dt (3 F_0^4 - 6 F_0^2 v^2 - v^4)\;. \label{Gamma1}
\end{eqnarray}

Repeating the procedure for the $F^6$ term, we obtain 

\begin{eqnarray}
  C_6 & = & \frac{F_0^6}{16}\left(J_1^6 + J_2^6 + J_3^6 - J_1^4J_2^2 
  - J_1^4J_3^2 - J_2^4J_3^2 - J_2^4J_1^2 - J_3^4J_1^2 - J_3^4J_2^2 
   \right. \nonumber \\
& &  \hspace{0.9cm} \left. + \, 2 J_1^2J_2^2J_3^2\right) \nonumber \\
& &    + \frac{F_0^4}{16}\left[J_0^2(J_1^4 + J_2^4 + J_3^4)
       -2J_0^2(J_1^2J_2^2 + J_1^2J_3^2 + J_2^2J_3^2)\right]v^2 \\
& & -\frac{F_0^2}{16}\left[J_0^4(J_1^2 + J_2^2 + J_3^2)\right]v^4 -
	\frac{1}{16}J_0^6 v^6\;. \nonumber 
\end{eqnarray}

A straightforward but lengthy calculation of the several traces 
in~\eqref{hc6} results in

\begin{eqnarray}
  \hat{C}_6 & = & -F_0^6
	\frac{N_6(n_6^2 + 4n_6 N_6 + 32 N_6^2)}{8(n_6 + N_6)}
   - F_0^4\frac{3n_6 N_6(n_6 + 3N_6)}{8(n_6 + N_6)} v^2 \nonumber \\
& &   - F_0^2 \frac{3n_6 N_6(n_6 + 2N_6)}{8 (n_6 + N_6)}v^4 - 
  \frac{n_6 N_6}{8}v^6\;. 
\end{eqnarray}

Reading from~\eqref{effsym} the 2-loop effective action, we have 

\begin{eqnarray}
   \Gamma^{(2)} & = & \frac{1}{2} \left(\frac{a_6}{M}\right)^2 
   	N \gymsq \int d^7x\;\hat{C}_6(F) \nonumber \\
  & = & \frac{g_s }{2^6(2\pi)^{7/2} r^2} V_6 \times \nonumber \\
  & & \int dt \left[-F_0^6 
	N_6(n_6^2 + 4n_6 N_6 + 32 N_6^2) 
	 - 3 F_0^4 n_6 N_6(n_6 + 3N_6)v^2 
		 \right. \nonumber \\
& &   \;\;\;\;\;\;\;\;\;\left. - 3 F_0^2 n_6 N_6(n_6 + 2N_6)v^4 - 
  N n_6 N_6 v^6 \right].
   \label{Gamma2}
\end{eqnarray}

%%%%%%%%%%%%%%%%%%%%%%%%%%%%%%%%%%%%%%%%%%%%%%%%%%%%%%%%%%%%%%%%%%%%%%

\section{Comparison}

We proceed to compare the supergravity and SYM results. In the two pictures 
there is a parameter that measures deviation from a supersymmetric 
state, $\epsilon$ in supergravity and $F_0$ in SYM. 
We expect them to be related.

In fact, from~\eqref{N0} we have 

\begin{equation}
  N_0 = \frac{F_0^3 N_6 V_6 }{(2\pi)^6}\;.
\end{equation}

Then~\eqref{MPQ} imply

\begin{equation}
  F_0^3 = \frac{Q}{P}\;,
\end{equation}

or, after replacing~\eqref{P},~\eqref{Q}, to leading order in $\epsilon$,

\begin{equation}
  F_0^2  =  \frac{\epsilon }{2\sqrt{3}}+ \frac{\epsilon^2}{12} + 
	O(\epsilon^{3})\;.
		\label{F02k} 
\end{equation}

So far we have used different 
normalisations in the supergravity and SYM calculations. In order to 
compare those results, we will from now on  express the 
SYM results in $\alpha^{\prime} = 1$  normalisation.

At 1-loop, restoring $T$ in~\eqref{Gamma1} and replacing 
the results~\eqref{MPQ}, \eqref{F02k}, 
we have, to leading order in $\epsilon$ for each order in $v$ and $1/r$,

\begin{equation}
 \Gamma^{(1)} =  \frac{n_6}{g_s (2\pi)^6}V_6 
	\int dt \left(\frac{\epsilon^2 M}{8r}
		-\frac{\epsilon \sqrt{3}M}{2r}v^2
		-\frac{M}{2r}v^4\right).
\end{equation}

Comparing with the supergravity result~\eqref{Ssugra06} we see that 
the SYM result 
reproduces exactly the $1/r$ and $v^2/r$ terms, a result already 
obtained in~\cite{bisy9711}.

At 2-loop level, 
repeating this procedure  
and considering the `source much heavier than probe' limit, 
$N_6 >> n_6$, we obtain, to leading order in $\epsilon$ in each 
order in $v$ and $1/r$,

\begin{equation}
 \Gamma^{(2)} =  \frac{n_6}{g_s (2\pi)^6}V_6 
	\int dt \left(-\frac{3 \epsilon^2 M^2}{4r^2}v^2
		-\frac{\sqrt{3}\epsilon M^2}{r^2}v^4
		-\frac{M^2}{r^2}v^6\right).
\end{equation}

Comparing with~\eqref{Ssugra06}, we see that the SYM result 
reproduces the $v^2/r^2$, $v^4/r^2$ and $v^6/r^2$ terms to leading order 
in $\epsilon$.

%%%%%%%%%%%%%%%%%%%%%%%%%%%%%%%%%%%%%%%%%%%%%%%%%%%%%%%%%%%%%%%%%%%%%

\section{Conclusion}

We studied the long distance, low velocity interaction potential 
between a D6-brane probe and a non-supersymmetric source containing D0- 
and D6-branes. We extended the supergravity 
 calculation of~\cite{bisy9711} to 
subleading order in $1/r$, and compared  the resulting potential 
with the one obtained using the ansatz of~\cite{che-tse9709} 
for the 2-loop SYM effective action. We found agreement at subleading 
order, thus providing a further non-trivial check for this ansatz.

It would be interesting to have a direct 
SYM calculation of the 2-loop terms, and to see how 
they compare to the ones obtained here.

\vskip 1cm \noindent
{\large{\bf {Acknowledgements}}}
\vskip 0.2cm
\noindent

We would like to thank Arkady Tseytlin for proposing 
the problem and for many useful discussions.

%%%%%%%%%%%%%%%%%%%%%%%%%%%%%%%%%%%%%%%%%%%%%%%%%%%%%%%%%%%%%%%%%%%%%%%%%%

\appendix

\section{Tr vs. tr}

This section is based in~\cite{che-tse9705,che-tse9709} and 
references therein.

Let Tr denote the trace in the adjoint representation, and tr denote 
trace in the fundamental representation. For $su(N)$ generators, 
$T_a$, and an element of the algebra, $X=X^a T_a$, the traces are 
related as 

\begin{eqnarray*}
\Tr(T_a T_b) & = & N \delta _{ab}\;, \\
\tr(T_a T_b) & = & \frac{1}{2} \delta_{ab}\;,\\
\Tr(X^2) & = & 2N\tr(X^2)\;,\\
\Tr(X^4)& = & 2N\tr(X^4)+6[\tr(X^2)]^2\;,\\
\Tr(X^6) & = & 2N\tr(X^6)+30\,\tr(X^4)\tr(X^2)-20[\tr(X^3)]^2\;.
\end{eqnarray*}

Similar relations apply to symmetrised products of generators,

\begin{eqnarray*}
\STr\left(X_{i_1}X_{i_2}X_{i_3}X_{i_4}\right) & = &
\Tr\!\left(X_{(i_1}X_{i_2}X_{i_3}X_{i_4)}\right)= \\
& = & 2N\tr\!\left(X_{(i_1}X_{i_2}X_{i_3}X_{i_4)}\right)+
	6\,\tr\!\left(X_{(i_1}X_{i_2}\right)
			\tr\!\left(X_{i_3}X_{i_4)}\right), \\
\STr\left(X_{i_1}X_{i_2}\ldots X_{i_6}\right) & = & 
\Tr\!\left(X_{(i_1}X_{i_2}\ldots X_{i_6)}\right) \\
& = & 2N\tr\!\left(X_{(i_1}X_{i_2}\ldots X_{i_6)}\right)+
	30\,\tr\!\left(X_{(i_1}\ldots X_{i_4}\right)
	\tr\!\left(X_{i_5}X_{i_6)}\right) \\
&& -20\,\tr\!\left(X_{(i_1}X_{i_2}X_{i_3}\right)
	\tr\!\left(X_{i_4}X_{i_5}X_{i_6)}\right).
\end{eqnarray*}

Note that even for commuting backgrounds, where obviously STr = Tr, 
it 
is often convenient to keep STr in order to change from Tr to tr.

We list some results that will be used. For arbitrary 
{\em commuting} $X_0$, $X_1$

\begin{displaymath}
  \STr\left(X_0^2 X_1^2\right) = 2N\tr\!\left(X_0^2 X_1^2\right) 
	+ 2\,\tr\!\left(X_0^2\right)\tr\!\left(X_1^2\right) + 
			4\,\left[\tr\left(X_0 X_1\right)\right]^2.
\end{displaymath}

Another common term is of the form $X_0^4 X_1^2$. Let $X$ represent 
either $X_0$ or $X_1$, such that $X^6 = X_0^4 X_1^2$. Then

\begin{eqnarray*}
sym\;\tr\!\left(X^4\right)\tr\!\left(X^2\right) 
	& = & \frac{1}{15}\left[\tr\!\left(X_0^4\right)\tr\!\left(X_1^2\right)+
	8\,\tr\!\left(X_0^3 X_1\right)\tr\!\left(X_0 X_1\right)\right. \\
	& &  \;\;\;\;\;\;\; +
	\left. 6\,\tr\!\left(X_0^2 X_1^2\right)
	\tr\!\left(X_0^2\right)\right], \\
sym\;\tr\!\left(X^3\right)\tr\!\left(X^3\right) 
	& = & \frac{1}{5}\left\{2\,\tr\!\left(X_0^3\right)
	\tr\!\left(X_0 X_1^2\right)+
	3\left[\tr\!\left(X_0^2 X_1\right)\right]^2\right\}, \\
	sym\;\tr\left(X^2\right)\tr\left(X^2\right)\tr\left(X^2\right) 
	& = & \frac{1}{5}
	\left\{\left[\tr\!\left(X_0^2\right)\right]^2\tr\!\left(X_1^2\right)+
	4\,\tr\!\left(X_0^2\right)
		\left[\tr\!\left(X_0 X_1\right)\right]^2\right\}, 
\end{eqnarray*}

where $sym$ denotes the symmetrisation operator.

Yet another common term is of the form $X_1^2 X_2^2 X_3^2$, all $X_i$'s 
commuting. 
Let $X$ represent one of the $X_i$, $i=1,2,3$, such that 
$X^6 = X_1^2 X_2^2 X_3^2$. Then

\begin{eqnarray*}
sym\;\tr\left(X^4\right)\tr\left(X^2\right) & = & \frac{1}{15}
   \left\{\left[\tr\!\left(X_1^2\right)
	\tr\!\left(X_2^2 X_3^2\right)+ \; 2 \; terms\right]\right. \\ 
       & &  \;\;\;\;\;\;\;
		\left. + \; 4 \left[\tr\!\left(X_1 X_2\right)
			\tr\!\left(X_1 X_2 X_3^2\right)
	 	+ 2 \; terms\right]\right\}, \\
sym\;\tr\left(X^3\right)\tr\left(X^3\right) & = & \frac{1}{5}
  \left\{2\left[\tr\!\left(X_1 X_2 X_3\right)\right]^2 +
	\left[\tr\!\left(X_1^2 X_2\right)
	\tr\!\left(X_2 X_3^2 \right) + 2 \; terms\right]\right\}, \\
sym\;\tr\left(X^2\right)\tr\left(X^2\right)\tr\left(X^2\right) 
	& = & \frac{1}{15}
	\left\{\tr\!\left(X_1^2\right)\tr\!\left(X_2^2\right)
		\tr\!\left(X_3^2\right) \right. \\
	& &  \;\;\;\;\;\;\;
	+ \; 2 \left[\tr\!\left(X_1^2 \right)\tr\!\left(X_2 X_3\right)
		\tr\!\left(X_2 X_3\right) + 2 \; terms\right]  \\ 
	& &  \;\;\;\;\;\;\;
	\left. + \; 8\,\tr\!\left(X_1 X_2\right)\tr\!\left(X_2 X_3\right)
		\tr\!\left(X_3 X_1\right)\right\}. 
\end{eqnarray*}

A common matrix appearing in these calculations is 

\begin{displaymath}
  J_0 = \frac{1}{N}\left(
   \begin{array}{cc}
       N_6 I_{n_6 \times n_6} &           0 \\
              0             &      -n_6 I_{N_6 \times N_6}
   \end{array} \right)
\end{displaymath}

where $N = n_6 + N_6\;$.

The adjoint trace of its even powers is given by 

\begin{displaymath}
\Tr(J_0^{2k}) = 2 n_6 N_6\;.
\end{displaymath}

%%%%%%%%%%%%%%%%%%%%%%%%%%%%%%%%%%%%%%%%%%%%%%%%%%%%%%%%%%%%%%%%%%%%%%

\end{document}